\documentclass[11pt]{article}
\usepackage[pdftex]{graphicx}
\usepackage{amsmath}
\usepackage{amssymb}

\begin{document}

\title{Biexciton as a Feshbach resonance and \\
Bose-Einstein condensation of paraexcitons in Cu$_{2}$O}

\author{Hoang Ngoc Cam \\
Institute of Physics, Vietnam Academy of Science and Technology, \\
10 Dao Tan, Ba Dinh, Hanoi, Vietnam \\}

\maketitle

\noindent Paraexcitons, the lowest energy exciton states in
Cu$_{2}$O, have been considered a good system for realizing exciton
Bose-Einstein condensation (BEC). The fact that their BEC has not
been attained so far is attributed to a collision-induced loss,
whose nature remains unclear. To understand collisional properties
of cold paraexcitons governing their BEC, we perform here a
microscopic consideration of the s-wave paraexciton-paraexciton
scattering. We show its two-channel character with incoming
paraexcitons coupled to a biexciton, which is a Feshbach resonance
producing a paraexciton loss and a diminution of their background
scattering length. The former elucidates the mechanism of the
long-observed paraexciton loss, which turns out to be inefficient at
temperatures near one Kelvin and below, whereas the latter makes the
paraexciton scattering length in strain-induced traps negative under
stress exceeding a critical value. Our rough estimates give this
value of order of one kilobar, hence already moderate stress creates
a serious obstacle to attaining a stable paraexciton BEC. Thus our
results indicate that BEC of trapped paraexcitons might be achieved
at a subkelvin temperature, but only under low stress.

\thispagestyle{empty}

\section*{Introduction}
Exciton in semiconductors is a Coulomb-bound pair of an electron in
the conduction band and a hole in the valence band. In the
low-density limit, excitons behave as bosons, so they may undergo
Bose-Einstein condensation (BEC) if their lifetime is long enough to
allow the system to reach quasiequilibrium.$^{1-3}$ Therefore BEC is
expected in Cu$_{2}$O where the dipole-forbidden 1s excitons of the
yellow series have relatively long lifetime. The state is split by
the electron-hole exchange into the singlet paraexciton and higher
lying triplet orthoexciton, separated by an energy of $\Delta =12$
meV.$^{4}$ The orthoexciton is quadruple allowed, while the
paraexciton is strictly forbidden resulting in its particularly long
lifetime.$^{5}$ Due to this property and yet to their large binding
energy, paraexcitons in Cu$_{2}$O have long been considered a good
system to realize exciton BEC in a three-dimensional solid.

Much effort has been made during the past several decades, but compelling
evidence of BEC in Cu$_{2}$O has not been obtained.$^{6-14}$ It is believed,
that the main obstacle is a collisional loss preventing paraexciton density
from achieving the critical value for BEC.$^{7,12-17}$ The loss was also
suggested to be the reason of the "explosion" observed by Gonokami's group
when they were able to realize the critical density for strain-confined
paraexcitons at a subkelvin temperature.$^{10}$ It is conventionally
attributed to Auger recombination, but its nature has not been elucidated.$%
^{18}$ Later, involvement of a biexciton$^{19,20}$ and inelastic collision
of paraexcitons$^{17}$ have been suggested, but their microscopic mechanism
is still open to make clear. In general, despite the recent experimental
progress toward paraexciton BEC,$^{10-13}$ a basic theoretical issue of the
problem -- the interparticle interaction, remains unsolved.

To understand and possibly to control obstacles to paraexciton BEC,
here we perform a microscopic consideration of the
paraexciton-paraexciton scattering at low temperatures. We begin by
formulating a Hamiltonian for the problem starting from its original
electron-hole picture. This enables us to establish the
interconversion between a pair of paraexcitons and that of
orthoexcitons, which results in the two-channel character of the
paraexciton scattering. To obtain salient features of the s-wave
collision dominating scattering at low temperatures, we develop an
approximate way of dealing with the nonlocal exchange part of
interaction potentials in two channels, which is also the coupling
potential. This makes it possible for us to estimate the paraexciton
background scattering length, the binding energy of a biexciton
supported by the closed channel as well as strength of the
paraexciton-biexciton coupling. With this coupling the biexciton is
not a bound state that can be detected, but a Feshbach resonance,
which manifests itself through changes it makes to collisional
properties of paraexcitons. Those include a paraexciton loss and an
extra attractive interaction added to their background repulsive
interaction. In strain-induced traps, wherein biexciton effects are
enhanced with stress, the paraexciton scattering length turns
negative as stress goes beyond a critical value $S_{0}$. Hence only
stress not higher than $S_{0}$, which in our estimates is of order
of one kilobar, can be used to create traps. This indicates that the
explosion reported in ref. 10 is connected with the negative
scattering length of strain-confined paraexcitons under moderate
stress. Concerning the loss, with a linear dependence on the
paraexciton-phonon scattering rate it can be made inefficient by
lowering temperatures to the range near one Kelvin and below. Our
results provide theoretical understanding of obstacles to
paraexciton BEC, which offers interpretation of recent experimental
results as well as suggests means to improve conditions for trapped
paraexcitons to reach their BEC.

\section*{Results}
\subsection*{Two-channel nature of paraexciton-paraexciton scattering}
We use second quantization formalism for an elucidation of the effective
spin-dependent two-body interaction among paraexcitons. We start from the
fact, that excitons are long-live quasiparticles introduced for an effective
description of electron-hole-pair systems\ with their inherent many-body
Coloumb-mediated correlations. Each exciton is "dressed" by the Coulomb
attraction of an electron with a hole and related to the other ones by the
"residual" interaction that comes from the remaining part of the many-body
correlations. Concerning the yellow-series 1s exciton in Cu$_{2}$O formed
from double degenerate $\Gamma _{6}^{+}$ \ and $\Gamma _{7}^{+}$ bands, it
splits into the triply degenerate orthoexciton $\Gamma _{5}^{+}$ with spin $%
J=1$ and nondegenerate paraexciton $\Gamma _{2}^{+}$ with $J=0$ according to
the group-theoretical expansion
\begin{equation}
\Gamma _{1}^{+}\otimes \Gamma _{6}^{+}\otimes \Gamma _{7}^{+}=\Gamma
_{2}^{+}\oplus \Gamma _{5}^{+},
\end{equation}%
where the unit representation $\Gamma _{1}^{+}$ characterizes symmetry of
the $1s$ hydrogenlike function describing the electron and hole relative
motion in the exciton. From the relationship between basis functions of
irreducible reperesentations $\Gamma _{2}^{+}$ and $\Gamma _{5}^{+}$ and
those of $\Gamma _{6}^{+}$\ and $\Gamma _{7}^{+}$, we have the momentum $%
\mathbf{k}$\ paraexciton state $P_{\mathbf{k}}^{+}|\,0\mathbf{\,)}$ and that
of orthoexcitons $O_{M\mathbf{k}}^{+}|\,0\mathbf{\,)}$ ($M=-1,0,1$ is the
orthoexciton spin projection on the quantization axis) expressed as
superpositions of correlated electron-hole pairs with total momentum $%
\mathbf{k}$.$^{21}$ With the aid of tables of Clebsch-Gordan coefficients
relevant to irreducible representations of the $O_{h}$ group,$^{22}$ we have%
\begin{equation}
P_{\mathbf{k}}^{+}\,|\,\mathbf{0\,)}=\frac{1}{\sqrt{V}}\sum_{\mathbf{p}%
}\digamma (\mathbf{p}-\beta \mathbf{k})\,\frac{1}{\sqrt{2}}\left( {\large {e}%
_{1/2,\mathbf{k}-\mathbf{p}}^{+}{h}_{-1/2,\mathbf{p}}^{+}+{e}_{-1/2,\mathbf{k%
}-\mathbf{p}}^{+}{h}_{1/2,\mathbf{p}}^{+}}\right) |\,0\,\rangle \,,
\end{equation}
\begin{equation}
O_{M\mathbf{k}}^{+}\,|\,\mathbf{0\,)}=\frac{1}{\sqrt{V}}\sum_{\mathbf{p}%
}\digamma (\mathbf{p}-\beta \mathbf{k})\times \left\{
\begin{array}{c}
{\large {e}_{1/2,\mathbf{k}-\mathbf{p}}^{+}{h}_{1/2,\mathbf{p}}^{+}}%
|\,0\,\rangle \,,\quad M=1, \\
\frac{1}{\sqrt{2}}\left( {\large {e}_{1/2,\mathbf{k}-\mathbf{p}}^{+}{h}%
_{-1/2,\mathbf{p}}^{+}-{e}_{-1/2,\mathbf{k}-\mathbf{p}}^{+}{h}_{1/2,\mathbf{p%
}}^{+}}\right) |\,0\,\rangle ,\quad M=0, \\
{\large {e}_{-1/2,\mathbf{k}-\mathbf{p}}^{+}{h}_{-1/2,\mathbf{p}}^{+}}%
|\,0\,\rangle \,,\quad M=-1,%
\end{array}%
\right.
\end{equation}%
where $|\,0\,\rangle $ is the semiconductor ground state and $|\,\mathbf{0\,)%
}$ -- that state mapped on the space of excitons, $V$ -- the sample volume, $%
\digamma (\mathbf{p}-\beta \mathbf{k})$ -- the $1s$ exciton envelope
function in the momentum space with $\beta =\mu _{h}/\mu _{x}$ the
hole-to-exciton mass ratio, $\mu _{x}=\mu _{e}+\mu _{h}$. Taking into
account the fact, that in relevant experiments only 1s excitons are excited,
we have the relationship between correlated electron-hole pairs and excitons
inverse to equations (2) and (3),%
\begin{eqnarray}
{\large {e}_{1/2,\mathbf{k}_{e}}^{+}\,{h}_{1/2,\mathbf{k}_{h}}^{+}}%
|\,0\,\rangle  &=&\frac{1}{\sqrt{V}}\digamma (\alpha \mathbf{k}_{h}-\beta
\mathbf{k}_{e})\,O_{1,\mathbf{k}_{e}+\mathbf{k}_{h}}^{+}|\,0\,),  \notag \\
{\large {e}_{1/2,\mathbf{k}_{e}}^{+}\,{h}_{-1/2,\mathbf{k}_{h}}^{+}}%
|\,0\,\rangle  &=&\frac{1}{\sqrt{V}}\digamma (\alpha \mathbf{k}_{h}-\beta
\mathbf{k}_{e})\,\left( P_{\mathbf{k}_{e}+\mathbf{k}_{h}}^{+}+O_{0,\mathbf{k}%
_{e}+\mathbf{k}_{h}}^{+}\right) |\,0\,),  \notag \\
{\large {e}_{-1/2,\mathbf{k}_{e}}^{+}\,{h}_{1/2,\mathbf{k}_{h}}^{+}}%
|\,0\,\rangle  &=&\frac{1}{\sqrt{V}}\digamma (\alpha \mathbf{k}_{h}-\beta
\mathbf{k}_{e})\,\left( P_{\mathbf{k}_{e}+\mathbf{k}_{h}}^{+}-O_{0,\mathbf{k}%
_{e}+\mathbf{k}_{h}}^{+}\right) |\,0\,),  \notag \\
{\large {e}_{-1/2,\mathbf{k}_{e}}^{+}\,{h}_{-1/2,\mathbf{k}_{h}}^{+}}%
|\,0\,\rangle  &=&\frac{1}{\sqrt{V}}\digamma (\alpha \mathbf{k}_{h}-\beta
\mathbf{k}_{e})\,O_{-1,\mathbf{k}_{e}+\mathbf{k}_{h}}^{+}|\,0\,).
\end{eqnarray}
\begin{figure}[t!!!]
\begin{center}
\includegraphics[width=0.54\textwidth]{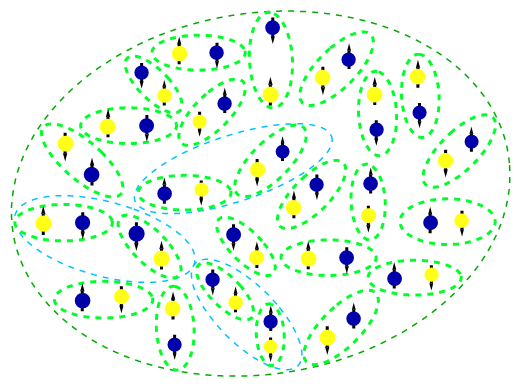}
\end{center}
\caption{\textbf{ Electron-hole representation of a system of paraexcitons}. Small
blue (yellow) balls with arrows depict electrons (holes) with spin-up and
spin-down. Dashed lightblue oval lines confine representative two-pair correlated
structures.}
\label{Figure 1}
\end{figure}

Thus, according to equation (2) a paraexciton system in the
electron-hole representation has the form of ensembles of zero-spin
correlated electron-hole pairs of the type schematically shown in
Fig. 1. The residual interaction among paraexcitons arises from
correlations between the electron and hole of each pair with all the
electrons and holes of the other pairs, which are governed by the
Coulomb forces and Pauli exclusion principle of
indistinguishability. It includes two-body, three-body, and so on
interactions, among which in dilute conditions the two-body
interaction dominates. There exist several ways of deriving an
exciton Hamiltonian with two-body effective interaction from
that of the original electron-hole system.$^{23-26}$ In our previous work$%
^{27}$ we have done this for the common case of direct-gap two-band
semiconductors by adopting the bosonization approach of
Hanamura$^{23,24}$ with the particles spin taken into consideration.
In the same way, here we map the correlations among the constituents
of two zero-spin correlated electron-hole pairs, which can be in
three possible spin combinations as shown by three examples in Fig.
1, onto the exciton space to obtain the Hamiltonian
\begin{eqnarray}
H_{p-p} &=&\sum\limits_{\mathbf{k}}E_{p}(k)P_{\mathbf{k}}^{+}P_{\mathbf{k}%
}+\sum\limits_{\mathbf{k}}E_{o}(k)\sum\limits_{M=-1,0,1}O_{M,\mathbf{k}%
}^{+}O_{M,\mathbf{k}}  \notag \\
&&+\frac{1}{2V}\sum\limits_{\mathbf{k}_{1},\mathbf{k}_{2},\mathbf{q}}\left\{
U^{d}(q)O_{1,\mathbf{k}_{1}+\mathbf{q}}^{+}O_{-1,\mathbf{k}_{2}-\mathbf{q}%
}^{+}O_{-1,\mathbf{k}_{2}}O_{1,\mathbf{k}_{1}}\right.  \notag \\
&&+\left[ U^{d}(q)+\frac{1}{2}U^{ex}(\mathbf{k}_{1}-\mathbf{k}_{2},\mathbf{q}%
)\right] \left[ P_{\mathbf{k}_{1}+\mathbf{q}}^{+}P_{\mathbf{k}_{2}-\mathbf{q}%
}^{+}P_{\mathbf{k}_{2}}P_{\mathbf{k}_{1}}+O_{0,\mathbf{k}_{1}+\mathbf{q}%
}^{+}O_{0,\mathbf{k}_{2}-\mathbf{q}}^{+}O_{0,\mathbf{k}_{2}}O_{0,\mathbf{k}%
_{1}}\right]  \notag \\
&&-\frac{1}{2}U^{ex}(\mathbf{k}_{1}-\mathbf{k}_{2},\mathbf{q})\left[ \left(
O_{1,\mathbf{k}_{1}+\mathbf{q}}^{+}O_{-1,\mathbf{k}_{2}-\mathbf{q}%
}^{+}+O_{-1,\mathbf{k}_{1}+\mathbf{q}}^{+}O_{1,\mathbf{k}_{2}-\mathbf{q}%
}^{+}\right) O_{0,\mathbf{k}_{2}}O_{0,\mathbf{k}_{1}}+h.c.\right]  \notag \\
&&\left. +\frac{1}{2}U^{ex}(\mathbf{k}_{1}-\mathbf{k}_{2},\mathbf{q})\left[
\left( O_{1,\mathbf{k}_{1}+\mathbf{q}}^{+}O_{-1,\mathbf{k}_{2}-\mathbf{q}%
}^{+}+O_{-1,\mathbf{k}_{1}+\mathbf{q}}^{+}O_{1,\mathbf{k}_{2}-\mathbf{q}%
}^{+}+O_{0,\mathbf{k}_{1}+\mathbf{q}}^{+}O_{0,\mathbf{k}_{2}-\mathbf{q}%
}^{+}\right) P_{\mathbf{k}_{2}}P_{\mathbf{k}_{1}} \right.\right. \notag \\
&&\left. \left. +h.c.\right] \right\} ,
\end{eqnarray}%
\begin{figure}[t!!!]
\begin{center}
\includegraphics[width=0.63\textwidth]{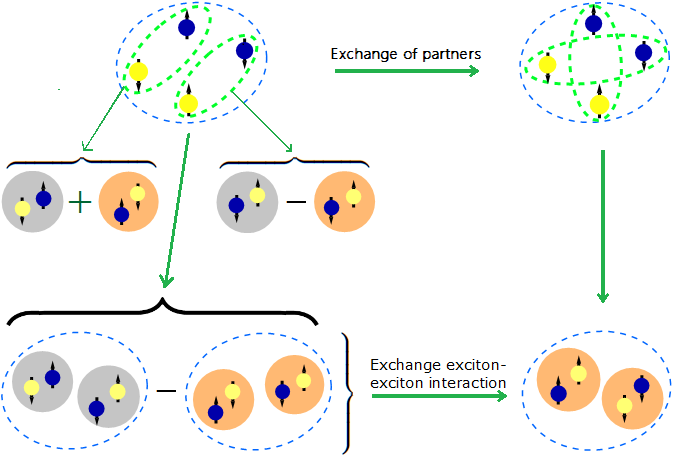}
\end{center}
\caption{\textbf{Mechanism of the exchange exciton-exciton interaction}.
Grey (orange) balls denote the dark paraexciton (bright
orthoexcitons) with the spin projection equal to the sum of those of the
constituent electron and hole. Left: On the top is one of complexes of two
correlated electron-hole pairs which constitute two correlated paraexcitons.
The complex includes two electrons (two holes) with opposite spins. Each
correlated pair is represented in the exciton space by a linear combination
of a paraexciton and a longitudinal orthoexciton in accordance with equation
(4). As a result, the whole complex is represented by a combination of one
correlated pair of paraexcitons and that of longitudinal orthoexcitons.
Right: The electron-hole complex on the top is obtained from the one on the
left by an exchange of partners between two pairs. It is represented in the
exciton space by a pair of correlated transverse orthoexcitons with opposite
spin directions. Thus, the exchange correlations between two electron-hole
pairs are reproduced in the exciton space by the exchange exciton-exciton
interaction. }
\label{Figure 2}
\end{figure}
where $E_{p}(k)$ and $E_{o}(k)$ denote respectively the paraexciton and
orthoexciton energy, $U^{d}$ and $U^{ex}$ -- energy densities of the direct
and exchange exciton-exciton interaction, which are functionals of $\digamma
$ depending parametrically on the mass ratios $\beta $ and $\alpha =1-\beta $%
,$^{23,26}$ and h.c. stands for hermitic conjugates. The last two terms in
braces show, that the exchange interaction can flip the spin of interacting
excitons. It either turns two longitudinal orthoexcitons ($M=0$) to a pair
of transverse ones ($M=1,-1$) with opposite spins, or convert a paraexciton
pair to a pair of orthoexcitons with zero total spin. Hereafter the term
orthoexcitons will be used to refer exclusively to these ones. As $U^{ex}>0$%
, we see that the two-boby interaction among paraexcitons is generally
repulsive except for the case of the exchange interaction between two
orthoexcitons, which is attractive. Mechanism of the exchange interaction is
schematically shown on an example in Fig. 2.

Thus, with the effective two-body interaction taken into consideration, a
"paraexciton system" incorporates not only paraexcitons, but also
orthoexcitons that go in pairs. In this connection, neither paraexciton
pairs nor orthoexciton ones can form eigenstates of Hamiltonian (5). They
form just components, or substates of eigenstates, which have the form of two-exciton vectors with definite momentum
\begin{eqnarray}
\Psi _{p-p}(\mathbf{K}) &=&\frac{1}{\sqrt{2V}}\sum\limits_{\mathbf{s}}\left[
\psi _{pp}(\mathbf{s})P_{\mathbf{s+K}/2\,}^{+}P_{\mathbf{-s+K/2}}^{+}+\frac{1}{\sqrt{3}}\psi _{po}(\mathbf{s})\left( O_{0,\mathbf{s+K}/2\,}^{+}O_{0,\mathbf{%
-s+K/2}}^{+} \right. \right.
\notag \\
&&\left. \left. O_{-1,\mathbf{s+K}%
/2\,}^{+}O_{1,\mathbf{-s+K/2}}^{+}+O_{1,\mathbf{s+K}/2\,}^{+}O_{-1,\mathbf{-s+K/2}}^{+}\right)
\right] \,\mid 0\mathbf{\,)},
\end{eqnarray}%
where $\psi _{pp}$ and $\psi _{po}$ are envelope functions respectively of
"bare" paraexciton and orthoexciton pairs. The Schrodinger equation $%
H_{p-p}\Psi _{p-p}(\mathbf{K})=E_{p-p}(k)\Psi _{p-p}(\mathbf{K})$ leads to\
a system of equations for $\psi _{pp}$ and $\psi _{po}$,
\begin{eqnarray}
-\left( \frac{\hbar ^{2}s^{2}}{\mu _{x}}+E\right) \psi _{pp}(\mathbf{s})+%
\frac{1}{V}\sum\limits_{\mathbf{q}}\left[ U^{d}(\mathbf{q})+\frac{1}{2}%
U^{ex}(2\mathbf{s},\mathbf{q})\right] \psi _{pp}(\mathbf{s}+\mathbf{q})
\notag \\
+\frac{\sqrt{3}}{2V}\sum\limits_{\mathbf{q}}U^{ex}(2\mathbf{s},\mathbf{q}%
)\psi _{po}(\mathbf{s}+\mathbf{q}) &=&0,  \notag \\
\left( -\frac{\hbar ^{2}s^{2}}{\mu _{x}}+2\Delta -E\right) \psi _{po}(%
\mathbf{s})+\frac{1}{V}\sum\limits_{\mathbf{q}}\left[ U^{d}(\mathbf{q})-%
\frac{1}{2}U^{ex}(2\mathbf{s},\mathbf{q})\right] \psi _{po}(\mathbf{s}+%
\mathbf{q})  \notag \\
+\frac{\sqrt{3}}{2V}\sum\limits_{\mathbf{q}}U^{ex}(2\mathbf{s},\mathbf{q}%
)\psi _{pp}(\mathbf{s}+\mathbf{q}) &=&0,
\end{eqnarray}%
where $\mu _{x}$ is assumed the same for both types of the exciton, $%
E=k_{B}T $ ($k_{B}$ is the Boltzmann constant, $T$ -- the temperature) is
paraexciton thermal energy with their scattering threshold chosen as the
energy zero. These equations describe two-channel paraexciton-paraexciton
scattering, wherein the open (background) channel of incoming paraexcitons
is coupled by the exchange interaction potential to the closed channel
formed by the interaction potential between two orthoexcitons having energy
higher than the paraexciton threshold energy by amount $2\Delta $. Hereafter
we confine ourselves to the s-wave scattering at such low temperatures, that
$E\ll 2\Delta $ and the rate of para-ortho up-conversion is practically
zero. In that case we can describe the scattering in the first approximation
by two uncoupled channels and take their coupling into consideration in the
next approximation.

\subsection*{Approximate interaction potentials and solutions for bare
channels}

To proceed, we need to define interaction potentials in bare channels. Their
direct part is known,$^{28}$ but we face the challenge of describing the
exchange part. As expected in the case, when real interactions in a
many-particle system are described in terms of an effective interaction
between quasiparticles, the exchange exciton-exciton interaction is
nonlocal. From a presentation for $U^{ex}$ (see Supplementary Information
(SI)) we observe, that the degree of its nonlocality decreases with decrease
of $\beta $ -- the small mass ratio, disappearing in the limit $\beta $ $%
\rightarrow 0$,\ when it equals the exchange energy in Heitler-London theory
of the hydrogen molecule.$^{29,30}$ This suggests, that the expansion of $%
U^{ex}$ into a series of powers of $\beta $ might provide a useful
approximation to the exchange interaction potential,%
\begin{eqnarray}
\frac{1}{2V}\sum\limits_{\mathbf{q}}U^{ex}(2\mathbf{s},\mathbf{q})\psi (%
\mathbf{s}+\mathbf{q}) &=&\int \exp [i\mathbf{sr}]d^{3}r  \notag \\
\times \left[ A_{0}(r)+\beta \,A_{1}(r)\frac{d}{dr}+\beta ^{2}\,A_{2}(r)%
\frac{d^{2}}{dr^{2}}+\beta ^{3}\,A_{3}(r)\frac{d^{3}}{dr^{3}}+...\right]
\phi (r) &\equiv &\int \exp [i\mathbf{sr}]d^{3}r\mathcal{V}^{ex}(r)\phi (r), \notag \\
&&
\end{eqnarray}%
if appropriately truncated. Here $\phi $ is any of s-wave functions of bare
channels in real space, and functions $A_{0}(r),A_{1}(r),...$ depend
parametrically on $\beta $ falling off exponentially at large distances. The truncation order depends on $\beta $ value, which
cannot be acquired from the band extremum value of electron and hole masses.
Because of the exciton considerable spread in momentum space$^{31}$ and
nonparabolicity of the valence band,$^{32}$ both $\mu _{h}$ and $\mu _{x}$
depend on the wave vector. We take the averaged value of $\beta $\ following
from the ratio $\mu _{r}/\mu _{x}=(1-\beta )\beta $ with reduced mass $\mu
_{r}$ drawn from the relation $E_{b}a_{x}^{2}=\hbar ^{2}/2\mu _{r}$ ($E_{b}$
and $a_{x}$\ are the exciton binding energy and effective radius,
respectively). Thus, to a set of values of the key parameters $E_{b}$, $a_{x}
$, and $\mu _{x}$, there corresponds a value of $\beta $ determining
interaction potentials. Let us take the commonly accepted values $a_{x}=0.7$
nm, $E_{b}=150$ meV and $\mu _{x}=2.6m_{0}$ ($m_{0}$ is the free electron
mass), then $\beta \approx 0.28$. For this value we can truncate expansion
(8) at the third term, and as a result, equations for the $\chi $-function ($%
\chi =r\,\phi /4\pi $) of the background s-wave scattering and possible
biexciton read%
\begin{equation}
2(\beta -1)\beta E_{b}\chi _{E}^{\prime \prime }+\left[ \mathcal{U}%
^{d}(x)+F_{0}(x)-E\right] \chi _{E}+\beta \,F_{1}(x)\chi _{E}^{\prime
}+\beta ^{2}\,F_{2}(x)\chi _{E}^{\prime \prime }+\beta ^{3}\,F_{3}(x)\chi
_{E}^{\prime \prime \prime }=0,
\end{equation}%
\begin{equation}
2(\beta -1)\beta E_{b}\chi _{b}^{\prime \prime }+\left[ \mathcal{U}%
^{d}(x)-F_{0}(x)+2\Delta -E_{0}\right] \chi _{b}-\beta \,F_{1}(x)\chi
_{b}^{\prime }-\beta ^{2}\,F_{2}(x)\chi _{b}^{\prime \prime }-\beta
^{3}\,F_{3}(x)\chi _{b}^{\prime \prime \prime }=0,
\end{equation}%
with $E_{0}$ the biexciton energy, $x\equiv r/a_{x}$ and functions $%
F_{0},F_{1},F_{2},F_{3}$ are expressed in terms of functions $%
A_{0},A_{1},A_{2},A_{3}$ (see SI). A change in deciding on a particular
value for a key parameter entails quantitative changes in solutions for bare
channels as well as their coupling, but the qualitative picture of the
paraexciton scattering remains the same as presented below for $\beta =0.28$.

Numerical solution of equation (9) for $E=0$ gives function $\chi _{0}$,
whose plot's intersection with the $x$-axis gives the background scattering
length $a_{bg}\approx 1.45a_{x}$. This result is considerably smaller than
our rough hard-core estimate$^{28}$ and that computed by quantum Monte Carlo
method for $\beta =\alpha \,$.$^{33}$ The quantity determines solely the
background phase shift $\delta _{bg}$ of slow paraexcitons.$^{21}$ Condition
for paraexcitons to be slow depends on the range of paraexciton-paraexciton
interaction potential $\mathcal{V}_{bg}=\mathcal{U}^{d}+\mathcal{V}^{ex}$,
which in its turn is defined by functions $\mathcal{U}^{d}(x)$ and $%
F_{0}(x),...,F_{3}(x)$. We get the range about $3.5a_{x}$, thus paraexcitons
having $k<<(3.5a_{x})^{-1}\approx 0.4$ nm$^{-1}$ are slow. To be specific,
we set the point $T=1.2$ K corresponding to $k\approx 0.06$ nm$^{-1}$ the
upper bound of the slow paraexcitons range.
\begin{figure}[t]
\begin{center}
\includegraphics[width=0.54\textwidth]{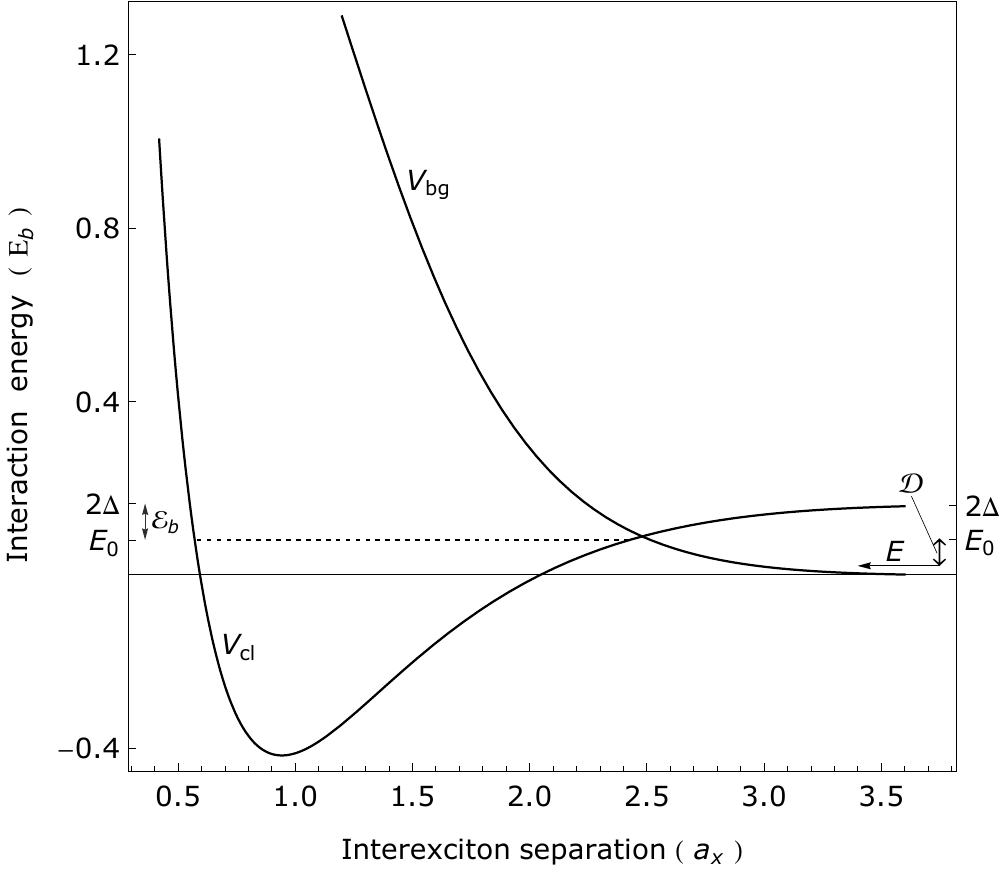}
\end{center}
\caption{\textbf{Approximate local equivalents of interaction potentials in
bare channels}. $V_{bg}$ is presented for $E=0$, whereas $V_{cl}$ -- for $%
E=E_{0}$. Their shape resembles that of the interaction potential between
hydrogen atoms respectively in their singlet and triplet molecular states.$%
^{21}$ For actual interexciton distances the interaction between
paraexcitons in the open channel is repulsive, whereas that between
orthoexcitons in the closed channel is attractive. With zero of energy
chosen at the paraexciton scattering threshold, the interaction potential
between orthoexcitons is shifted upward by the energy of $2\Delta $
involving the biexciton energy (dotted line) to be "embedded" in the
scattering continuum.}
\label{Figure 3}
\end{figure}
As to equation (10), we find by the customary variational procedure, that it
has one bound state -- the biexciton with binding energy $\mathcal{E}%
_{b}=2\Delta -E_{0}\approx 12.78$ meV and the corresponding wave function%
\begin{equation}
\chi _{b}=1.22267\,x\,\exp [-1.8\exp (-2.475\,x)][\exp (-2.48\,x)]^{0.279}.
\end{equation}%
The obtained value for $\mathcal{E}_{b}$ is much larger than the result of
Brinkman et al.$^{34}$ and almost coincides with that of Huang.$^{35}$

To have an idea of the shape of potentials $\mathcal{V}_{bg}$ and $\mathcal{V%
}_{cl}=\mathcal{U}^{d}-\mathcal{V}^{ex}$, we perform localization procedures
to obtain their approximate local equivalents.$^{36}$ Neglecting for
simplicity small $\beta ^{3}$-terms in equations (9) and (10), we turn them
into usual Schrodinger equations by transformations $\chi
_{E}(x)=T_{E}(x)\Phi _{E}(x)$ and $\chi _{b}(x)=T_{b}(x)\Phi _{b}(x)$,
respectively. The acquired energy-dependent local potentials $V_{bg}$ and $%
V_{cl}$ are shown in Fig. 3.

\subsection*{Effects of biexciton as a Feshbach resonance}

The coupling of two channels, when it is "turned on", induces oscillations
between paraexciton scattering states and the biexciton, which is no longer
a bound state, but just a quasistationary state,$^{21}$ or a scattering
resonance$^{37}$ with an energy uncertainty $\Gamma _{c}$ and the
corresponding lifetime $\tau =\hbar /\Gamma _{c}$. This explains the fact
why biexciton has not been detected in Cu$_{2}$O. In quantum-mechanical
scattering theory, the resonance width, which is called also the coupling
strength, is obtained from the second-order correction to the discrete
energy level caused by perturbation in the form of the coupling potential.$%
^{21,37}$ In our case, $\Gamma _{c}\propto \int \sqrt{E}dE\,\left[ \langle
\phi _{E}|\mathcal{V}^{ex}|\phi _{b}\rangle \right] ^{2}\delta (E-E_{0})=$ $%
\sqrt{E_{0}}\,\left[ \langle \phi _{E_{0}}|\mathcal{V}^{ex}|\phi _{b}\rangle %
\right] ^{2}$. To get the transition matrix element, we solve equations (7)
in the real space for $\phi _{pp}=\phi _{E_{0}}$ and $\phi _{po}=\phi _{b}$
with the approximate $\mathcal{V}^{ex}$, taking into account the fact that $%
\phi _{b}=4\pi \chi _{b}/r$ is the solution of equation (10).
Computations give $\Gamma _{c}\approx 7.85$ meV, which is of the
same order as the biexciton binding energy. Resonance scattering of
paraexcitons happens when their energy matches the interval of width
$\Gamma _{c}$ around $E_{0}$ through the mechanism illustrated in
Fig. 4. Such a phenomenon is long known in nuclear physics as a
Feshbach resonance.$^{38}$ They have lately become an important
experimental tool for controlling properties of cold atomic
gases.$^{39}$
\begin{figure}[t!!!]
\begin{center}
\includegraphics[width=0.45\textwidth]{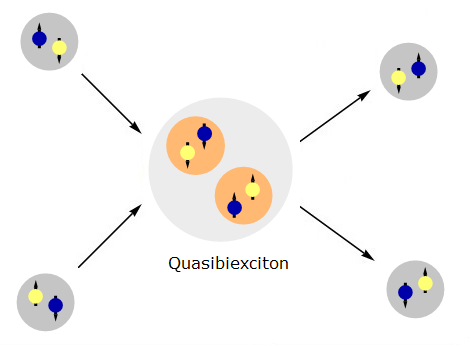}
\end{center}
\caption{\textbf{Resonance scattering of paraexcitons on the quasibiexciton}%
. Two incoming paraexcitons encounter, interact via spin-flip exchange
interaction potential and convert to an orthoexciton pair, which attracts
each other forming temporarily the quasibiexciton. After the time $\protect%
\tau $ elapses, two orthoexcitons convert to a pair of paraexcitons leading
to the quasibiexciton decay.}
\label{Figure 4}
\end{figure}
In comparison with magnetic Feshbach resonances in the last, the
quasibiexciton in Cu$_{2}$O has two distinctive features. First, its strong
coupling with the open channel by the exchange exciton-exciton interaction
potential, which is at once the ruling part of interaction potentials in
bare channels. In this connection the resonance effects hold over a wide
off-resonance area, in contrast to atomic gases, where they take place in a
narrow resonance area. Second, the quasibiexciton is a decaying state by
itself having a width $\Gamma _{qb}$ connected with all damping mechanisms
-- partly a consequence of that, the system under consideration is not quite
closed, but coupled to surroundings.

Mixing of paraexcitons with the quasibiexciton as a decaying resonance gives
rise to a resonance factor in their s-wave scattering matrix element,$%
^{21,37}$%
\begin{equation}
S_{0}=\exp [2i\delta _{bg}]\left[ 1+i\frac{\Gamma _{c}}{\mathcal{D}-i(\Gamma
_{c}+\Gamma _{qb})/2}\right] ,
\end{equation}%
where $\mathcal{D}$ is the detuning of scattering energy from the resonance,
$\mathcal{D}=2\Delta -\mathcal{E}_{b}-E$ (see Fig. 3). From here we have the
total s-wave scattering cross section $\sigma _{t}=2\pi (1-\mathrm{Re}%
S_{0})/k^{2}$,
\begin{equation}
\sigma _{t}=\frac{\pi }{k^{2}}\left[ 4\sin ^{2}\delta _{bg}+\frac{\Gamma
_{c}(\Gamma _{c}+\Gamma _{qb})}{\mathcal{D}^{2}+(\Gamma _{c}/2+\Gamma
_{qb}/2)^{2}}\cos (2\delta _{bg})+2\frac{\mathcal{D}\Gamma _{c}}{\mathcal{D}%
^{2}+(\Gamma _{c}/2+\Gamma _{qb}/2)^{2}}\sin 2\delta _{bg}\right] .
\end{equation}%
As seen, coupling to the quasibiexciton causes a resonance scattering
superimposed on the paraexciton background scattering. This is described by
the second and last terms in brackets presenting respectively the resonance
scattering and its interference with background scattering. They include two
quasibiexciton effects. First, a paraexciton loss connected with an
inelastic component in the resonance scattering, whose rate $A=v\sigma _{r}$
($v$ is the mean velocity of the particle undergoing scattering) is defined
by the inelastic cross section $\sigma _{r}=\pi (1-|S_{0}|^{2})/k^{2}$,
\begin{equation}
A=4\pi \frac{\hbar ^{2}}{\mu _{x}}\sqrt{\frac{6}{\mu _{x}E}}\frac{(\Gamma
_{c}/2)(\Gamma _{qb}/2)}{\mathcal{D}^{2}+(\Gamma _{c}/2+\Gamma _{qb}/2)^{2}}.
\end{equation}%
Second, a change in the paraexciton-paraexciton interaction described by the
elastic cross section $\sigma _{e}=\sigma _{t}-\sigma _{r}$. Obviously, the
terms other than 4$\sin ^{2}\delta _{bg}$ in $\sigma _{e}$ together
represent an additonal interaction joining up with the background
paraexciton-paraexciton interaction. Rich physics of resonance effects,
whose scale is defined by the relation between the quasibiexciton total
width and detuning, is beyond the scope of this paper. Here we consider them
in the off-resonance area, where $(\Gamma _{c}/2+\Gamma _{qb}/2)/\mathcal{D}%
<1$, which is pertinent to conditions of recent experiments. Then, being of $%
(\Gamma _{c}/2\mathcal{D})^{2}$\ scale, the resonance term in $\sigma _{e}$
is relatively small. Hence we have for slow paraexcitons,
\begin{equation}
\sigma _{e}\left\vert _{ka_{bg}<<1}\right. \approx 4\pi \left\{ a_{bg}-\frac{%
1}{k}\frac{\Gamma _{c}}{2\mathcal{D}}\left[ 1+\left( \frac{\Gamma _{c}}{2%
\mathcal{D}}+\frac{\Gamma _{qb}}{2\mathcal{D}}\right) ^{2}\right]
^{-1}\right\} ^{2}.
\end{equation}%
Thus, coupling to the quasibiexciton gives rise to an attractive interaction
characterized by the term $a_{res}\propto -\Gamma _{c}/2\mathcal{D}$ in the
paraexciton scattering length $a=a_{bg}+a_{res}$. Regarding $A$, along with $%
\Gamma _{c}/2\mathcal{D}$, it is proportional to the ratio $\Gamma _{qb}/2%
\mathcal{D}$. Let us assume, that the quasibiexciton decay width is about
two times that of paraexcitons, $\Gamma _{qb}/2\approx \Gamma _{p}$. Besides
a negligibly small population relaxation rate, $\Gamma _{p}$ comprises
homogeneous broadening due to the exciton-exciton and exciton-phonon
scattering, $\Gamma _{x-x}$ and $\Gamma _{ph}$, and inhomogeneous broadening
$\Gamma _{inh}$ due to fluctuations of the sample structure and applied
fields. Consequently, the loss rate depends on the density, temperature,
sample quality as well as inhomogeinity of external fields over the sample.
We see, that $\Gamma _{ph}$ makes the loss persistent at low densities$%
^{7,17}$ and $\Gamma _{inh}$ is a source of uncertainty causing measurements
of the loss rate diverge.$^{14}$

\subsection*{Quantitative estimates and interpretation of recent
experimental results}

\begin{figure}[t!!!]
\begin{center}
\includegraphics[width=0.54\textwidth]{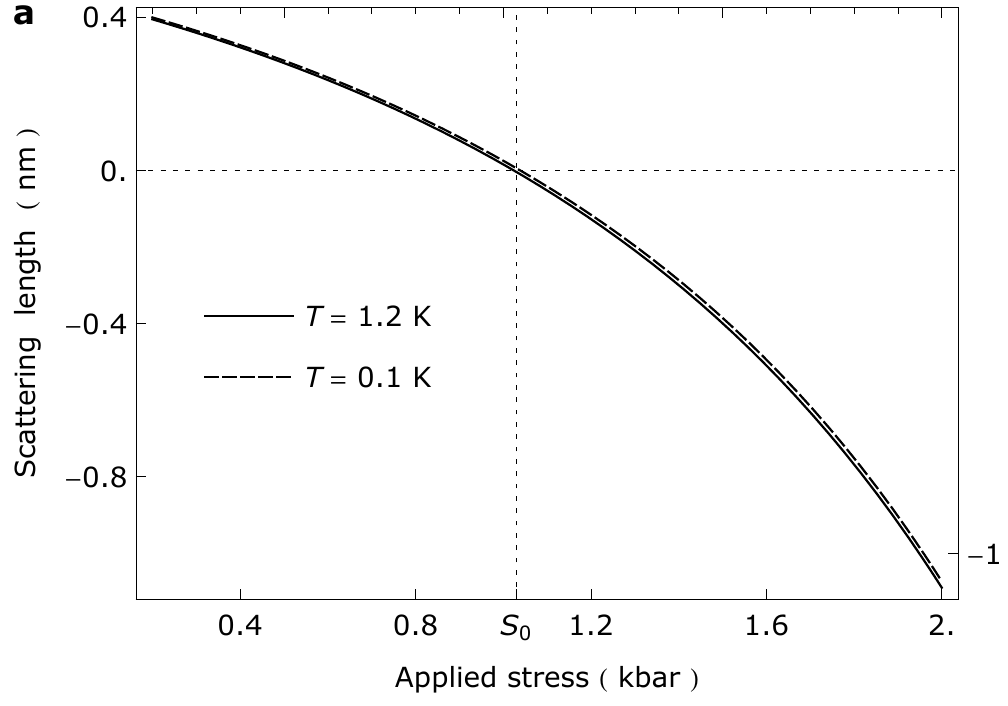}
\par
\includegraphics[width=0.477\textwidth]{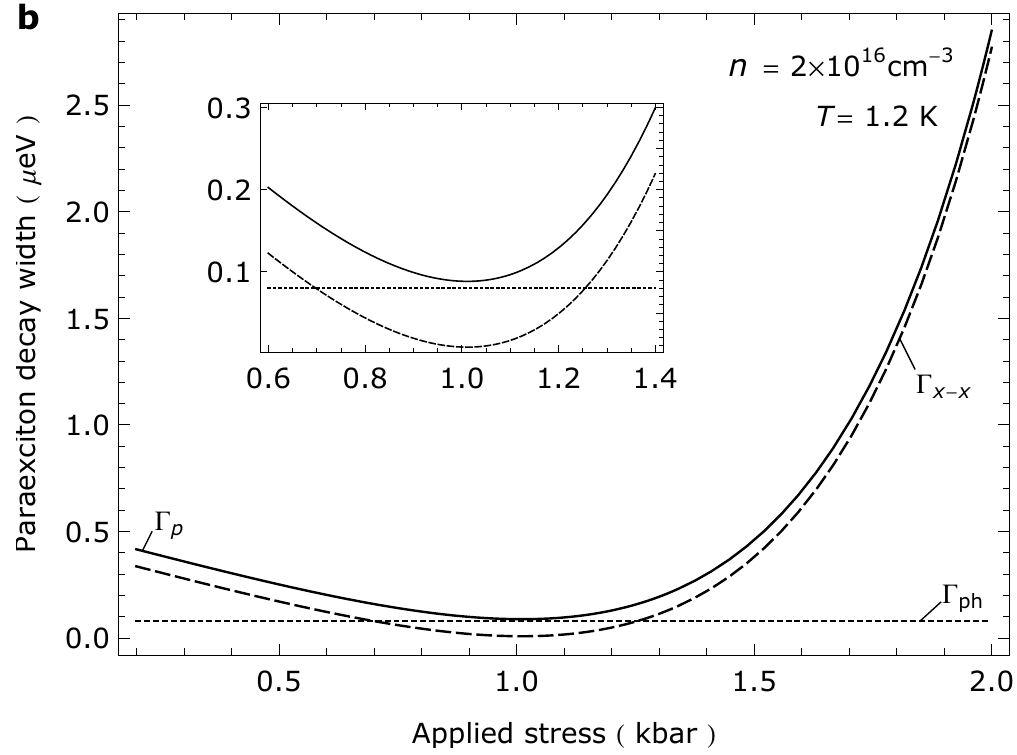} %
\includegraphics[width=0.471\textwidth]{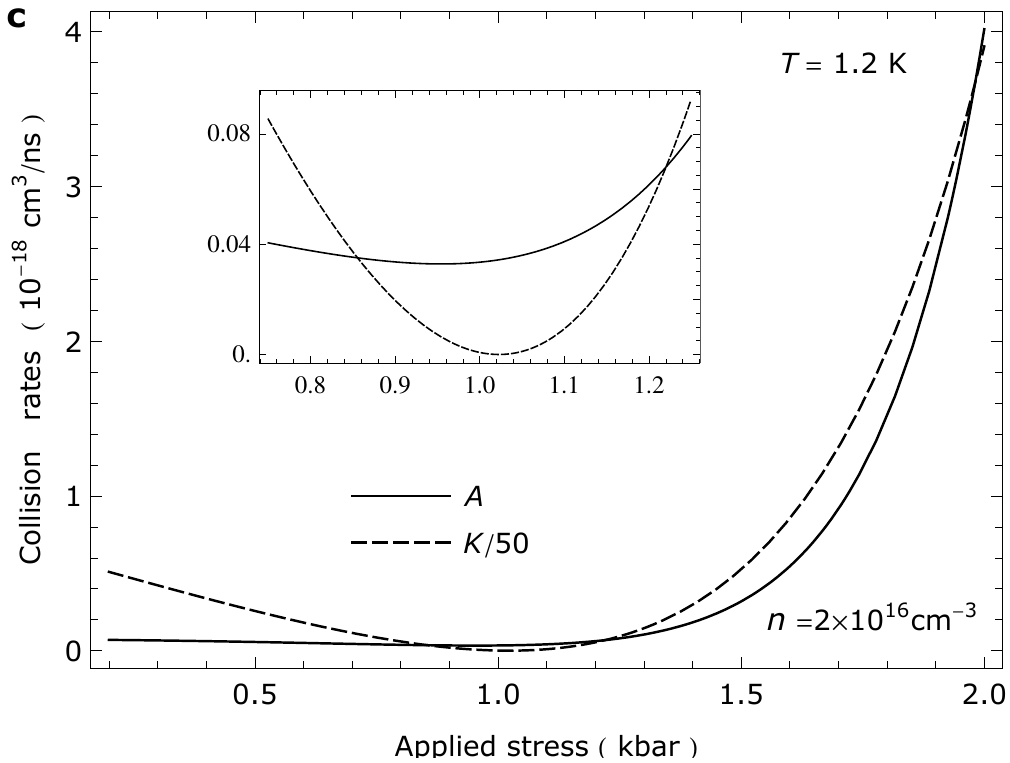}
\end{center}
\caption{\textbf{Stress dependence of the parameters describing
quasibiexciton effects on scattering of slow paraexcitons}. (\textbf{a})
Scattering length decreases monotonically with the applied stress changing
its sign at $S_{0}\approx 1$ kbar, then rapidly gains considerable negative
values: under $S=2$ kbar, $\left\vert a\right\vert \approx 1.1$ nm $>$ $%
a_{bg}$. The results depend little on the temperature. (\textbf{b})
Paraexciton decay width and its two components in the absence of
inhomogeneous broadening. Being nearly proportional to $a^{2}$, at small
values of the applied stress $\Gamma _{x-x}$ decreases moderately taking its
minimum at $S_{0}$, then rises speeding up when $S$ approaches 2 kbar. So
does $\Gamma _{p}$, which differs from $\Gamma _{x-x}$ by the small constant
$\Gamma _{ph}$. The relationship between $\Gamma _{p},\Gamma _{x-x}$ and $%
\Gamma _{ph}$ in the interval round 1 kbar is shown magnified in the inset. (%
\textbf{c}) Paraexciton loss rate on the background of the fiftyfold reduced
elastic collision rate $K$. The behaviour of $K$ is almost the same as $%
\Gamma _{x-x}$, but unlike the latter, it does not depend on the density. As
to $A$, it varies slightly in the range before 1.1 kbar, but after the point
grows steeply due to both increase of $\Gamma _{p}$ and decrease of $%
\mathcal{D}$ (see equation (15)). The inset shows the relationship between $A
$ and reduced $K$ in the vicinity of their minimum.}
\label{Figure 5}
\end{figure}
We limit our rough estimates to scattering of slow paraexcitons in
harmonic traps induced by moderate uniaxial stress -- conditions
covering those of recent experiments on BEC in Cu$_{2}$O.$^{10-13}$
Such traps are necessary to avoid paraexciton heating and diffusion
as well as to lower the critical density.$^{7,12}$ By reducing
$\Delta $,$^{40,41}$ stress moves the quasibiexciton position
downwards the scattering threshold (see Fig. 3). Coupling of
scattering states to a resonance close to the threshold is
energy-dependent. As a consequence, one has to replace the coupling
strength $\Gamma _{c}$ in equations (12) -- (15) by the function
$\gamma _{r}(E)=\Gamma _{c}\sqrt{E}\diagup (E_{0}^{2}+\Gamma
_{c}^{2}/4)^{1/4}.^{21}$ From stress dependence of $\Delta $ we find
that the off-resonance condition
corresponds to the applied stress $S\lesssim 2$ kbar. At $T\leq 1.2$ K $%
\gamma _{r}(E)/\mathcal{D}$ is small and $\Gamma _{p}/\mathcal{D}$ --
negligibly small (we see later that $\Gamma _{p}$ is at most several
microelectronvolt). Accordingly, equations (14) and (15) give the measure of
quasibiexciton effects in the approximate form
\begin{figure}[t!!!]
\begin{center}
\includegraphics[width=0.474\textwidth]{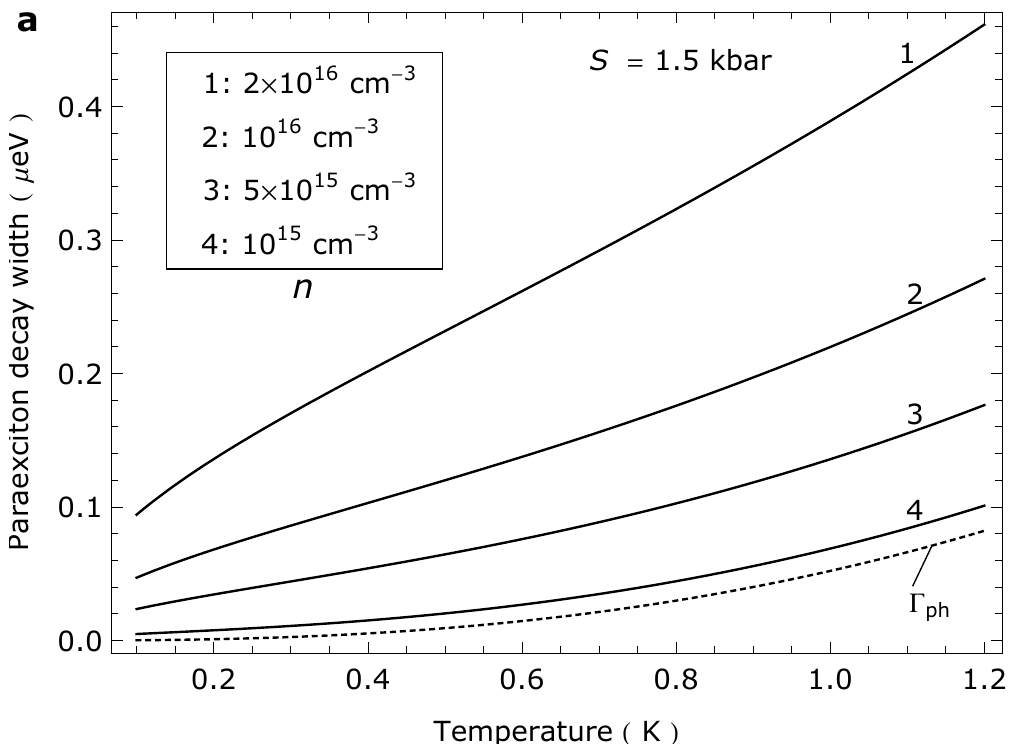} %
\includegraphics[width=0.474\textwidth]{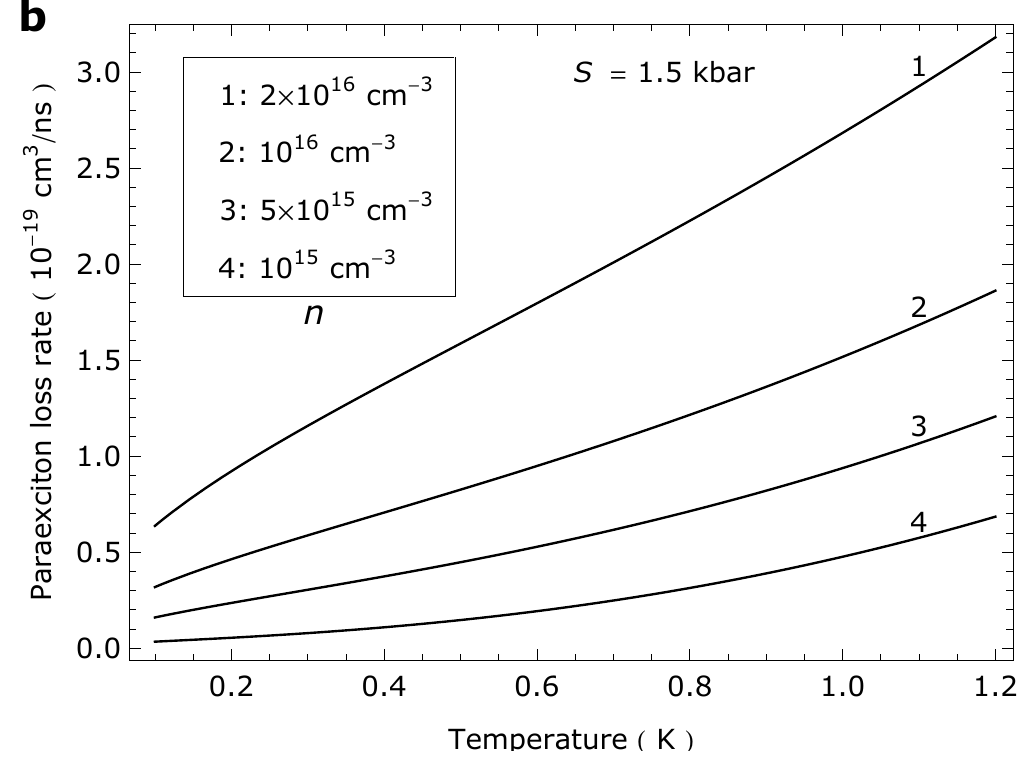}
\end{center}
\caption{\textbf{Temperature dependence of parameters characterizing the
decay and loss of slow paraexcitons at different densities}. (\textbf{a})
Paraexciton decay width varying with temperature under applied stress of 1.5
kbar as $\Gamma _{p}\simeq (0.052T^{5/2}+0.015\mathfrak{n}T^{1/2})$ $\protect%
\mu $eV/K ($\mathfrak{n}$ -- density measured in $10^{15}$ cm$^{-3}$). The
exciton-exciton scattering contributes to the decay already at a density as
low as $10^{15}$ cm$^{-3}$. (\textbf{b}) Paraexciton loss rate decreasing
with temperature nearly in the same way, as its decay width. For the density
$2\times 10^{16}$ cm$^{-3}$ realized in today's standard experiments, the
loss rate falls off threefold when temperature drops from 1.2 K to 0.2 K.}
\label{Figure 6}
\end{figure}
\begin{equation}
a_{res}\propto -\sqrt{\frac{\hbar ^{2}}{\mu _{x}}}\frac{\Gamma _{c}}{2%
\mathcal{D}},\;\,A\propto \frac{\hbar ^{2}}{\mu _{x}}\frac{\Gamma _{c}}{2%
\mathcal{D}}\frac{\Gamma _{p}}{\mathcal{D}},
\end{equation}%
showing, that $a$, and with it the elastic collision rate $K=\sigma
_{e}v\propto a^{2}$, are independent of $\Gamma _{p}$. On contrast,
the loss rate $A\propto $ $\Gamma _{p}$, which at low densities and
small inhomogeneity equals $\Gamma _{ph}$. Because interaction of
excitons with longitudinal acoustic phonons is stress-independent
and that with transverse acoustic ones is weak,$^{42}$ we assume the
value of\ $\Gamma _{ph}$\ at 1.2 K under moderate stress to be the
same 80 neV as measured under zero stress.$^{43}$ Neglecting in the
following inhomogeneity of the
strain field, we put $\Gamma _{p}\simeq \Gamma _{x-x}+\Gamma _{ph}$ with $%
\Gamma _{x-x}=\hbar n\,(A+K)$ computed by iterations and obtain $A$ for
different $S$ at a particular density. Along with $a$, which decreases with
increasing $S$ (see Fig. 5a), these parameters are sensitive to $S$ (see
Fig. 5b \& 5c). Including a term proportional to $a^{2}$, they have a
minimum at $S_{0}$, where $a=0$, then begin to rise gaining steep increase
when stress approaches 2 kbar, which is close to the resonance area of $%
\Gamma _{c}/2\mathcal{D}\gtrsim 1$. Actually, the increase of $A$ with
stress has been observed at moderate and high stress values.$^{16}$ Figure
5b shows, that apart from an interval around $S_{0}$, where $\Gamma
_{ph}>\Gamma _{x-x}$, in general under stress the paraexciton decay is
governed by exciton-exciton scattering. Concerning the elastic and inelastic
collision rates, the former absolutely dominates with the ratio $K/A$
ranging from about 20 near $S_{0}$ to over 50 away from the point (see Fig.
5c). The factor is favorable for evaporative cooling of thermal paraexciton
clouds.$^{44}$ The domination holds with decreasing temperature, because $%
K\propto T^{1/2}$ and $A$\ drops slightly faster. In fact, as seen from Fig.
5b, for $S\approx 1$ kbar, $\Gamma _{ph}\simeq \Gamma _{p}$, so we get from
temperature dependence of paraexciton mobility under $S=1$ kbar\ $\Gamma
_{ph}(T)\propto T^{5/2}$.$^{41}$ As to $\Gamma _{x-x}$, for densities up to $%
10^{17}$ cm$^{-3}$, $\Gamma _{x-x}\simeq \hbar n\,K\propto nT^{1/2}$. Thus $%
A\propto \Gamma _{p}=$ $(c_{1}nT^{1/2}+c_{2}T^{5/2})$, where $c_{1}$, $c_{2}$
-- constants. The two parameters are plotted against temperature in Fig. 6.
Figure 6a indicates another time, that under an applied stress outside
vicinity of $S_{0}$, the exciton-exciton elastic scattering plays the
leading part in paraexciton decay down to $n=5\times 10^{15}$cm$^{-3}$. At $%
n=2\times 10^{16}$ cm$^{-3}$, the loss rate values at $0.8$ K and $0.2$ K
drawn from Fig. 6b are about $2\times 10^{-19}$\ cm$^{3}$/ns and $10^{-19}$\
cm$^{3}$/ns, respectively, which are an order of magnitude less than those
reported in refs. 12 and 13. Presumably, inhomogeneity of the strain field
introduces some addition.
\begin{figure}[t!!!]
\begin{center}
\includegraphics[width=0.54\textwidth]{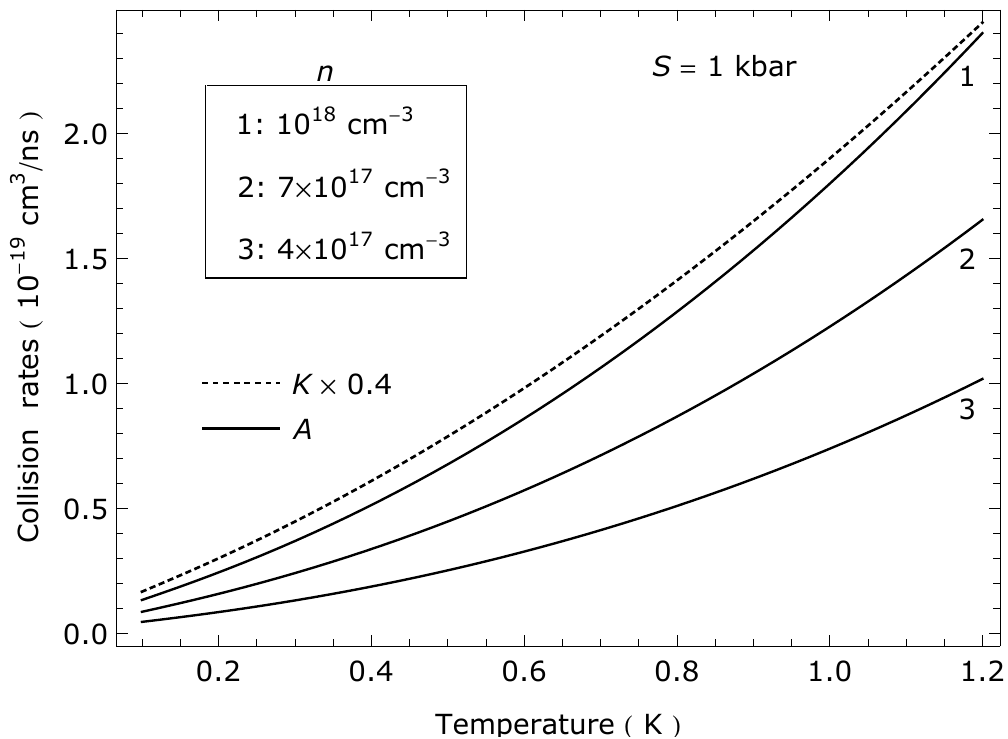}
\end{center}
\caption{\textbf{Temperature dependence of the paraexciton elastic collision
rate and loss rate at high densities}. Taking the trap from ref. 10 with $%
2\times 10^{9}$ paraexcitons, the density of $10^{18}$ cm$^{-3}$, $7\times
10^{17}$ cm$^{-3}$ and $4\times 10^{17}$ cm$^{-3}$ corresponds to the
condensation fraction 12\%, 5\% and 1\%, respectively.}
\label{Figure 7}
\end{figure}

Returning to Fig. 5a, one sees that for $S>S_{0}$ paraexcitons have $a<0$
corresponding to an effective attractive interaction. BEC of a trapped Bose
gas with $a<0$ is possible,$^{45-47}$ but unstable leading to the
condensate's collapse.$^{48-50}$ This happens when the number of condensate
particles $N_{0}$ exceeds a critic number $N_{cr}\approx 0.46\,a_{ho}/|a|$ ($%
a_{ho}$ -- the mean harmonic oscillator length). For the trap with $%
a_{ho}\approx 0.615$ $\mu $m under $S=1.4$ kbar (corresponding to $a\approx
-0.295$ nm) from ref. 10, $N_{cr}\approx 960$. Probably, that this number
was exceeded in the condensate the authors achieved, so it collapsed and
subsequently exploded. To avoid this, stress $S\leq S_{0}$ ensuring
nonnegative $a$ must be used. Let us consider the limit of strong
interaction $N_{0}a/a_{ho}\gg 1$, wherein the condensate radius increases
with $N_{0}$, $R\approx a_{ho}(15N_{0}\,a/a_{ho})^{1/5}$.$^{45}$ It is
reasonable to take a value of $S$ in the vicinity of $S_{0}$, when the decay
and loss parameters have their minimum, say 1 kbar corresponding to $%
a\approx 0.018$ nm. Then for the trap from ref. 10, $a/a_{ho}\approx 3\times
10^{-5}$. Let $N_{0}$ be as high as $2\times 10^{8}$ yielding $%
N_{0}\,a/a_{ho}\approx 6\times 10^{3}$ and $R\approx 6$ $\mu $m, which
corresponds to $n=N_{0}/R^{3}\approx 9\times 10^{17}$ cm$^{-3}$. From Fig. 7
we find, that at 0.8 K and $n=10^{18}$ cm$^{-3}$, $A\approx 1.3\times
10^{-19}$ cm$^{3}$/ns. Inhomogeneous broadening might somewhat raise the
value, but explosion due to small $a$ is excluded. We note in passing, that
domination of the elastic collision rate over loss rate remains even for
such a high density. Thus, a stable paraexciton condensate might take place
under experimental conditions of ref. 10 provided the applied stress $S\leq
S_{0}$.

\section*{Discussion}

It is the analogy of excitons with atoms that has motivated the search for
BEC in Cu$_{2}$O. However, the distinguishing feature is that excitons are
composite entities made from charge-carriers -- fermions, whose
Coulomb-mediated correlations are governed by the Pauli exclusion principle.
The internal structure of excitons produces the spin-dependent
exciton-exciton interaction that governs their BEC. In the case of
yellow-series 1s excitons in Cu$_{2}$O, it results in the interconversion
between a pair of paraexcitons with mutual repulsive interaction and that of
orthoexcitons, which attract each other, leading to the two-channel
character of the paraexciton-paraexciton scattering. We have described the
scattering at low temperature in the approximation of coupled bare channels
accepted in scattering theory. By an approximate way of calculating the
nonlocal exchange exciton-exciton interaction potential, we have been able
to estimate the paraexciton background scattering length $a_{bg}$, binding
energy $\mathcal{E}_{b}$ of a biexciton supported by the closed channel as
well as strength $\Gamma _{c}$ of the coupling between the biexciton and
paraexciton scattering states. The coupling makes the biexciton a Feshbach
resonance with its two characteristic effects -- a loss of particles in the
open channels and a diminution of their background scattering length, which
give us clues about the mechanism of obstacles to paraexciton BEC. First,
the loss of paraexcitons observed in numerous experiments is elucidated,
which turns out to be connected with continued conversion of two
paraexcitons into the quasibiexciton as an intermediate state. Reflecting
the paraexciton decaying nature characteristic of semiconductor electronic
excitations, the loss rate is proportional to the paraexciton decay width,
which comprises of the exciton-exciton and exciton-phonon scattering widths
and also inhomogeneous broadening. The last explains divergence of
measurements of the loss rate in different experiments. At relatively high
temperatures, when exciton-phonon scattering is effective and rises with
temperature, the rate can be high. In a unstressed crystal, it is likely the
main reason of the paraexciton saturation at high densities. However, of our
interest in this paper are strain-confined paraexcitons at temperatures near
one Kelvin and below. By shifting paraexciton and orthoexciton energies in
different ways, stress reduces the ortho-para splitting and by this enhances
quasibiexciton effects. We have found that, although the component related
to the exciton-exciton scattering increases with stress, steep decrease of
exciton-phonon scattering rate at such low temperatures makes the
paraexciton loss inefficient to affect BEC. Collisional properties of
paraexcitons is then determined solely by their scattering length $a$, which
is affected by the other quasibiexciton effect. At an applied stress $S_{0}$%
, the background scattering length is balanced by the diminution induced by
the quasibiexciton, so $a=0$, then it turns negative for $S>S_{0}$. Our
rough numerical estimates give $S_{0}$ of order of one kilobar. Thus we have
revealed that already moderate stress, which is generally used to produce
potential traps for paraexcitons to prevent their diffusion as well as to
lower their critical density, makes the paraexciton scattering length
negative. As BEC of trapped bosons with $a<0$ is unstable leading to the
condensate's collapse, this creates\ a serious obstacle to attaining a
convincing BEC. This result is supported by the process of `explosion' of
paraexcitons under stress of 1.4 kbar below the critical temperature seen by
Gonokami's group.

Our quantitative estimates, in particular that of the critical value $S_{0}$
of applied stress, have relied on three parameters $a_{bg}$, $\mathcal{E}%
_{b} $ and $\Gamma _{c}$ computed by using approximate exchange interaction
potential. The potential shape is sensitive to $\beta $, which is rather
uncertain in connection with some discrepancy in measurements of exciton key
parameters $E_{b},a_{x}$ and $\mu _{x}$. A slight modification of $\beta $
can lead to considerable variations of $\mathcal{E}_{b}$ and $\Gamma _{c}$,
and consequently, changes the value of $S_{0}$. But that is not the whole
picture. The approximation of two coupled bare channels applied in this
paper is a good one for scattering energy $E\ll 2\Delta $. However, its
disadvantage lies in the estimate of the coupling strength $\Gamma _{c}$,
which has been obtained from the second-order correction to the
quasibiexciton energy produced by the coupling potential as a perturbation.
That was a rough approximation because the exchange exciton-exciton
interaction potential is not weak. The perturbation caused by this potential
to quasistates of bare channels seems more complicated. In this context, our
treatment gives just general qualitative features of collisional properties
of cold paraexcitons leaving much room for quantitative improvement by
experiments as well as by other theoretical approaches.

\section*{Acknowledgements}

The author would like to thank M. Kuwata-Gonokami and K. Yoshioka
for valuable discussions, wherein initial ideas for this work were
formed, and S. A. Moskalenko and Nguyen Thanh Phuc for critical
reading of the manuscript. The financial support from Vietnam
National Foundation for Science and Technology Development
(NAFOSTED) through Grant No. 103.01-2014.73 is acknowledged.

\end{document}